\documentclass[]{article}

\usepackage[sc]{mathpazo} 
\usepackage[T1]{fontenc} 
\usepackage{microtype} 

\usepackage[english]{babel} 

\usepackage[hmarginratio=1:1,top=32mm,columnsep=20pt]{geometry} 
\usepackage[hang, small,labelfont=bf,up,textfont=it,up]{caption} 
\usepackage{booktabs} 

\usepackage{lettrine} 

\usepackage{enumitem} 
\setlist[itemize]{noitemsep} 

\usepackage{abstract} 

\usepackage{titlesec} 
\titleformat{\section}[block]{\large\scshape\centering}{\thesection.}{1em}{} 
\titleformat{\subsection}[block]{\large}{\thesubsection.}{1em}{} 

\usepackage{fancyhdr} 
\usepackage{titling} 

\usepackage{hyperref} 

\usepackage{epsfig}
\usepackage{graphicx}	
\usepackage{amsmath}	
\usepackage{amssymb}	
\usepackage{float}
\usepackage{subcaption}

\pretitle{\begin{center}\Large\bfseries} 
	\posttitle{\end{center}} 
\title{Development of a model to investigate the effect of the bias in SNIa measurements related to the inhomogeneity of space} 
\author{%
	Vincent Deledicque \\[1ex] 
	\normalsize \textit{No affiliation} \\ 
	\normalsize \href{mailto:vincent.deledicq@gmail.com}{vincent.deledicq@gmail.com} 
}
\date{\today} 
\renewcommand{\maketitlehookd}{%
	\begin{abstract}
		\noindent It has been suggested recently that the appparent accelerated expansion of the universe could be explained by a bias in the SNIa measurements. Such events indeed occur mainly in overdense regions, where matter is located, and whose dynamics can perhaps not been considered as representative of the one of the universe. In this article, we develop a model to investigate in more detail the effect of this bias. This model depends on one single parameter, related to the void fraction of space, and leads to simple analytical relations. We in particular determine the average metric tensor in overdense regions, and deduce that the scale factor and the rate at which time progresses in such regions differ significantly from the corresponding values expected on the average space. We then quantitatively deduce how redshift and luminosity distance measurements are affected by the bias, taking into account the perturbation of the metric tensor. Using a value for the void fraction corresponding to the order of magnitude found in the literature, we show that the model is able to predict a distance modulus versus redshift relation being in excellent aggreement with the one corresponding to a universe characterized by $\Omega_{m,0} = 0.3$ and $\Omega_{\Lambda,0} = 0.7$.
	\end{abstract}

}

\begin{document}
	
	\maketitle
	
	
	\section{Introduction}	
		
	The Cosmological principle, according to which space is considered as homogeneous and isotropic at a global scale, is an assumption that provides an enormous simplification in the investigation of the dynamics of the universe. A consequence of this assumption is that space is a maximally symmetric manifold, leading to the well-known Friedmann-Lema\^\i tre-Robertson-Walker (FLRW) metric. This is one of the simplest approaches that can be followed to model a dynamic universe, since it allows to study the universe with one single evolving parameter, namely the scale factor $a$.
	
	On the basis of this model, results obtained from several cosmological probes have led to conclude that the expansion of the universe was accelerating since some time. Such results were first obtained by \cite{Riess} and \cite{Perlmutter} from the observations of distant Type Ia supernovae (SNIa), and were supported by the additional measurements performed over the years, see \cite{Scolnic}. Similar results were also obtained using other independent cosmological probes, based in particular on cosmic microwave background (see in particular the WMAP \cite{Bennett} and the Planck projects \cite{Aghanim}) or baryon acoustic oscillations (BAO) measurements (see \cite{Eisenstein}).
	
	Despite a relatively broad consensus in the scientific community on the reality of the accelerated expansion, generally attributed to some form of dark energy, alternative theories have been proposed to explain the results obtained by those cosmological probes. Recently, it has been suggested by \cite{Deledicque} that a bias in the SNIa measurements could result in an apparent dark energy effect, explaining thus the accelerated expansion as an observational artefact. The bias is related to the fact that space is not homogeneous, as supposed in theory, and that SNIa occur preferentially in overdense regions, in which matter has grouped together. Since those regions have their own dynamics, which cannot be considered as representative of the one of the universe at large scale, measurements performed on SNIa should necessarily be altered by this inhomogeneity. Its was shown in $\cite{Deledicque}$ by using a global approach that this bias could explain the appearance of an apparent cosmological constant in the Einstein equation of General Relativity. However, this global approach did not allow for a detailed understanding of the underlying cause in the SNIa measurements that led to observe an accelerated expansion. The aim of this article is to continue the work of $\cite{Deledicque}$, by developing a model to better understand how SNIa measurements could be affected by the bias.
	
	In section $\ref{S1}$ we develop a two-regions model, assuming that space is made of overdense and underdense regions, both being characterized by their own average metric and stress-energy tensors. We examine in particular the dynamics of overdense regions, and show that it significantly diverges from the one of the global universe, characterized by the FLRW metric. In section $\ref{S2}$, we then use this model to investigate the effect of the inhomogeneity on the measurements performed on SNIa, in particular on luminosity distance and redshift measurements. In section $\ref{S3}$, we finally present and discuss the results of the model. We show that the bias introduced by the inhomogeneity in space leads to observe an apparent accelerated expansion, and that the predicted distance modulus versus redshift relation that would be measured taking into account the bias is in excellent agreement with the one reported by \cite{Riess}, \cite{Perlmutter} or \cite{Scolnic}. 
	
	
	\section{The two-regions model}\label{S1}
	
	From now on, we will admit that the cosmological constant vanishes, so the Einstein equation of General relativity reads
	\begin{equation}\label{GR}
		G_{\mu\nu} = 8\pi G T_{\mu\nu}\,.
	\end{equation}
	Using this equation together with the FRLW metric, assuming a flat space, we find that the scale factor $a$ presents a dynamics verifying the Friedman equation:
	\begin{equation}\label{FLRW}
		3\frac{\dot{a}^2}{a^2} = 8\pi G \rho\,,
	\end{equation}
	where $\rho$ is the average density over space, and where $a$ has been normalized to be dimensionless, such that $a = 1$ at the current time.
	 
	This approach, however, does not allow to investigate the effect of the bias identified in $\cite{Deledicque}$. Indeed, if initially matter was distributed in an almost homogeneous way, small perturbations developed, sharpening progressively the local inhomogeneous character of space. Due to the gravitational attraction, matter grouped together in regions having larger densities than on average, leading consequently to regions of void that expanded over time. SNIa occur preferentially in regions where matter is present, hence in overdense regions. Since they do not occur randomly over space, but instead in specific regions only, which probably cannot be considered as representative of the universe, this has to be considered as a bias if they are used as a cosmic probe. 
		
	We develop in this section the simplified model that will be used in our investigation of the effect of the bias on SNIa measurements. The simplest approach that can be followed is to consider space to be made of two different kinds of regions: overdense regions and underdense regions. To fix ideas, let us consider a volume $V$ of space sufficiently large so that it can be assumed to be representative of the universe. The average density of matter in $V$ is written as $\rho$, and corresponds to the density used in the Friedman equation. In this volume $V$, overdense regions occupy a volume $V_o$ and have an average density $\rho_o$, while underdense regions occupy a volume $V_u$ and have an average density $\rho_u$. We will consider here the limiting case for which $\rho_u = 0$, hence underdense regions do not contain any matter. The total mass in $V$ is $M = \rho V$, but since all matter is assumed to be located in overdense regions, we also have $M = \rho_o V_o$. We thus deduce that
	\begin{equation}\label{ak}
		\frac{V}{V_o} = \frac{\rho_o}{\rho}\,.
	\end{equation}
		
	Overdense and underdense regions can have extremely complicated characteristics, but we do not want to examine them in all their complexity. In order to simplify as much as possible the analysis, we will consider that overdense (resp. underdense) regions can be characterized by one single typical region, assumed to representative of the average behaviour of all existing overdense (resp. underdense) regions. Moreover, we are not interested in having a detailed knowledge of the spatial evolution of the metric or of the stress-energy tensor through space in this typical region, we will therefore admit that typical regions can be described by average tensors. Such an approach is also followed when using the FLRW metric, but here we apply it at a smaller scale. So, for our model we have two typical regions, one for the overdense regions, and one for the underdense regions, both having their own average metric and average stress-energy tensors.
	
	Let us thus consider such a typical region supposed to be representative of the average of all overdense regions. To describe the metric of this region, we will use a specific frame of reference based on the co-moving coordinates, for which $t$ is the cosmological time coordinate, and where $(x,y,z)$ are the spatial Cartesian coordinates. 
	
	Since space is assumed to be globally isotropic, we will admit that this is also the case for this typical region. The region supposed to be representative of the average of all existing overdense regions should indeed not present any directional preference. In the considered frame of reference, the metric tensor for a typical region can then be written as
	\begin{equation}\label{az}
		g_{\mu\nu} = \left(
		\begin{array}{c c c c}
			-f^2 & 0 & 0 & 0\\
			0 & b^2  & 0 & 0\\
			0 & 0 & b^2 & 0\\
			0 & 0 & 0 & b^2
		\end{array} \right)\,.
	\end{equation}
	Here, $b$ is the scale factor inside the typical region, equivalent to the scale factor $a$ for the global universe. Since the metric is considered to be the spatial average over the typical overdense region, $b$ does not depend on spatial coordinates. It however may depend on $t$. We notice also that the first diagonal component is not necessarily equal to $-1$, as for the FLRW metric. Indeed, the rate at which time evolves in a typical overdense regions will in general differ from the one in the FLRW metric, and hence it is characterized by a function $f$ depending on $t$ at most.

	For this metric, the Christoffel symbols are such that $\Gamma^{0}_{\ 00} = \dot{f}/f$, $\Gamma^{0}_{\ ii} = b\dot{b}/f^2$, $\Gamma^{i}_{\ it} = \Gamma^{i}_{\ ti} = \dot{b}/b$ and all other components are zero. In these relations, we use the dot notation to represent a derivative with respect to the cosmological time. Then, the Ricci tensor components are
	\begin{eqnarray}
		R_{00} &=& 3\left(\frac{\dot{f}\dot{b}}{fb} - \frac{\ddot{b}}{b}\right)\,,
		\\
		R_{ii} &=& \frac{2\dot{b}^2 + b\ddot{b}}{f^2} - \frac{b\dot{b}\dot{f}}{f^3}\,,
	\end{eqnarray}
	and all other components are zero. So, the Ricci scalar is	
	\begin{equation}
		R = 6\left(\frac{\ddot{b}}{bf^2} + \frac{\dot{b}^2}{b^2f^2} - \frac{\dot{b}\dot{f}}{bf^3}\right)\,.
	\end{equation}
	Finally, the components of the Einstein tensor are
	\begin{eqnarray}
		G_{00} &=& 3\frac{\dot{b}^2}{b^2}\,,
		\\
		G_{ii} &=& 2\frac{b\dot{b}\dot{f}}{f^3} - 2\frac{b\ddot{b}}{f^2} - \frac{\dot{b}^2}{f^2}\,,
	\end{eqnarray}
	and all other components are zero.
	
	The typical overdense region is assumed to be globally at rest in the comoving coordinates. As a consequence, its average four-velocity is
	\begin{equation}
		U^\mu = \frac{dx^\mu}{d\tau} = \left(\frac{1}{f}, 0, 0, 0\right)\,.
	\end{equation}	
	Matter being supposed to behave as a perfect fluid, the stress-energy tensor in a typical overdense region, defined as
	\begin{equation}
		T_{\mu\nu} = \left(\rho_o + p_o\right)U_\mu U_\nu + p_0 g_{\mu\nu}\,,
	\end{equation}
	becomes
	\begin{equation}\label{aq}
		T_{\mu\nu} = \left(
		\begin{array}{c c c c}
			\rho_o f^2 & 0 & 0 & 0\\
			0 & p_ob^2  & 0 & 0\\
			0 & 0 & p_ob^2 & 0\\
			0 & 0 & 0 & p_ob^2
		\end{array} \right)\,,
	\end{equation}
	where $p_o$ is the pressure in the typical overdense region. We then notice that the conservation law $\nabla_\mu T^\mu_{\ 0} = 0$ leads to
	\begin{equation}\label{ee}
		\dot{\rho_o} = -3\frac{\dot{b}}{b}\left(\rho_o + p_o\right)\,.
	\end{equation}
	
	In a typical overdense region, we will admit that matter has quite rapidly reached a state in which it is gravitationally bound. A region made of gravitationally bound matter has a constant volume over time. Normally, it should grow due to the expansion of the universe, but this growth is compensated by the gravitational attraction. The volume being constant, and the region containing a given mass, this means that the density in overdense regions remains constant over time. Hence, according to Eq.\ $(\ref{ee})$, overdense regions have a pressure $p_{o} = -\rho_{o}$. 
		
	Applying the equation of General Relativity Eq.\ $(\ref{GR})$ to our typical overdense region, using the relations obtained above, we find for the first diagonal component
	\begin{equation}\label{E1}
		3\frac{\dot{b}^2}{b^2} = 8\pi G\rho_{o} f^2\,,
	\end{equation}
	while for the three other diagonal components we have
	\begin{equation}\label{E2}
		2\frac{b\dot{b}\dot{f}}{f^3} - 2\frac{b\ddot{b}}{f^2} - \frac{\dot{b}^2}{f^2} = 8\pi G p_{o}b^2\,.
	\end{equation}
	As for the Friedman equations, we can show that Eq.\ $(\ref{E2})$ can be obtained from Eq.\ $(\ref{E1})$, its first temporal derivative and the fact that $p_{o} = -\rho_{o}$. We will thus only examine Eq.\ $(\ref{E1})$.
	
	Eq.\ $(\ref{E1})$ expresses the dynamics of the scale factor $b$ inside the typical overdense region, in function of its density $\rho_o$ and the function $f$. Since the right hand side of $(\ref{E1})$ is not zero, we expect that $b$ will not remain constant over time. In a globally expanding universe, $b$ will grow over time, as does $a$, but with a different rate. But if $b$ increases, this could mean that the volume of the considered typical region increases as well. On the other hand, we have considered that matter was gravitationally bound in overdense regions, so the volume of overdense regions should not change. In fact, while $b$ increases and involves a swelling of the overdense region, simultaneously, matter moves back to keep constant the volume it occupies. It is this phenomenon that is responsible for the negative pressure in this region.
	
	The approach that has been followed so far for a typical overdense region can be followed in a similar way for a typical underdense region. On average, such a region should also be isotropic, and the average scale factor for that region will verify a similar relation as Eq.\ $(\ref{E1})$, in which however the right hand side term is zero, because of the absence of matter in underdense regions. This means that Eq.\ $(\ref{E1})$ predicts a constant scale factor for a typical underdense region. This does not mean, however, that underdense regions will not grow over time. Indeed, as explained above, even if typical overdense regions keep a constant volume, their scale factor increases. The related volume increase is compensated by a backward displacement of matter to keep the volume of the overdense region constant. In moving back, matter leaves hence some void volume that initially belonged to an overdense regions, but that will contribute to enlarge the underdense region. So, underdense regions grow by a continuous volume transfer coming from the overdense regions.
	
	Let us quantity this volume transfer. At a given time, the volume of overdense regions is proportional to $b^3$. If matter was unbound, the rate at which this volume would grow is proportional to $3b^2\dot{b}$, with the same constant of proportionality. Per unit volume, this rate is thus equal to $3\dot{b}/b$. So, in the considered volume $V$ of the universe, the total rate of growth of overdense regions if matter was unbound is equal to $3V_o\dot{b}/b$. But since matter is assumed to be gravitationally bound, overdense regions are not expected to grow, and the increase of volume just calculated is supposed to be the one that will be transferred to underdense regions. In fact, as this volume increase is the only one that occurs in the volume $V$ (due to a null density, underdense regions do not grow by themselves), it should also correspond to the volume increase of the whole volume $V$. In other words, the rate at which overdense regions increase (if matter was not gravitationally bound) corresponds to the rate at which the universe expands. Given that the volume $V$ is expected to obey on average the Friedman equation, its rate of growth should be equal to $3V\dot{a}/a$. Hence, we have:
	\begin{equation}\label{fq}
		3V_o\frac{\dot{b}}{b} = 3V\frac{\dot{a}}{a}\,.
	\end{equation}
	Combining this latter equation with Eq.\ $(\ref{FLRW})$ and $(\ref{E1})$, we find that
	\begin{equation}
		f = \frac{V}{V_o}\sqrt{\frac{\rho}{\rho_o}}\,.
	\end{equation}
	Using Eq.\ $(\ref{ak})$, this can also be written as
	\begin{equation}\label{a1}
		f = \sqrt{\frac{\rho_o}{\rho}} = \sqrt{\frac{V}{V_o}} > 1\,.
	\end{equation}
	Since $V$ is proportional to $a^3$ and $V_o$ is assumed to be constant, we deduce that $f$ is proportional to $a^{3/2}$. Clearly, $f$ may significantly differ from $1$, meaning that the first diagonal component of the metric tensor in overdense regions diverges from the one of the FLRW metric, which is representative of the global universe on average. In overdense regions, time progresses at a rate larger than the one on average through space.
	
	It is interesting to notice that in a flat matter dominated universe, $a$ is proportional to $t^{2/3}$, implying thus that $f$ is directly proportional to $t$. In other words, $f$ linearly increases with the cosmological time:
	\begin{equation}\label{fqw}
		f = f_0t\,,
	\end{equation}
	where $f_0$ is a constant to determine.	Knowing the temporal dependence of $f$, we can integrate Eq.\ $(\ref{E1})$. We find
	\begin{equation}
		b = b_0\exp\left(\sqrt{\frac{2\pi}{3}G\rho_{o}}f_0 t^2\right)\,,
	\end{equation}
	where $b_0$ is another constant to determine.
		
	To fix $f_0$ and $b_0$, it is first important to highlight the limitations of the model. Eq.\ $(\ref{fqw})$ seems to show that $f$ will tend to zero for smaller times, whereas we would expect that $f$ should tend to $1$. The metric of overdense regions should indeed tend to the FLRW metric when going back in time, because perturbations were much smaller and space was looking to be more and more homogeneous. The explanation is as follows. We have considered that the volume $V_o$ of overdense regions is constant over time. This assumption makes sense as long as the total volume $V$ is larger than $V_{o}$. Obviously the volume of overdense regions must be included in the total volume. But $V$ being proportional to $a^3$, going back in time, at some point, $V$ will become smaller than $V_0$. So at times at which $V < V_{o}$, our model does not hold anymore. In fact, we notice from Eq.\ $(\ref{a1})$ that when $V = V_{o}$, thus when the overdense region occupies whole space, $f = 1$. This is the starting point of our model. At that point, the overdense region coincides with whole space, and we expect that its metric is equivalent to the FLRW metric, thus we should also have $b=a$. At earlier times, we cannot assume that $V_o$ remains constant, it should on the contrary shrink at the rate of $a^3$, and it will continuously coincide with the whole existing volume $V$. So, the constants of proportionality $f_0$ and $b_0$ are such that at the time when $V = V_o$, we have $f = 1$ and $b=a$.
	
	Let us calculate which value of $a^*$ can be considered as the starting point of our model. As just said, it should be such that $f = 1$. Knowing that $V_o$ is constant and that $V$ is proportional to $a^3$, we write
	\begin{equation}
		\frac{V}{V_o} = Aa^3\,,
	\end{equation}
	where $A$ is a constant equal to the current value of $V/V_o$, given that by convention at the current time $a=1$. As we will see through this article, $A$ is an important parameter, since it is the single one that completely fixes the whole model. So, from Eq.\ $(\ref{a1})$, we deduce that
	\begin{equation}\label{ff}
		f = \sqrt{A}a^{3/2}\,,
	\end{equation}
	and hence that $a^* = A^{-1/3}$.
	
	Eq.\ $(\ref{ff})$ provides a relation of $f$ as a function of $a$. For practical reasons, it will also be useful to have a relation of $b$ as a function of $a$. From Eq.\ $(\ref{fq})$ we deduce that
	\begin{equation}
		\frac{db}{da} = \frac{V}{V_o}\frac{b}{a} = Aa^2b\,.
	\end{equation}
	Integrating this last relation, we get
	\begin{equation}
		b = B\exp\left(\frac{A}{3}a^3\right)\,.
	\end{equation}
	The constant $B$ can be found by imposing that $b=a$ when $a=a^*$. We find
	\begin{equation}
		B = \left(A\right)^{-1/3}\exp\left(-\frac{1}{3}\right)\,.
	\end{equation}
	We have established the temporal dependence of $f$ and $b$, so the metric of a typical overdense region is completely determined.	
		
	
	\section{Effect of the bias on the SNIa measurements}\label{S2}
		
	Having established the metric tensor in a typical overdense region, we will now investigate the effect of the bias in SNIa measurements. This requires to understand how the inhomogeneity of space affects redshift and luminosity distance measurements.
	
	\subsection{Effect on redshift measurements}
	
	Redshift measurements allow to deduce the scale factor at the point where the SNIa occurred, simply by performing some specific time span measurement at our current epoch. To fix ideas, let us consider a source (typically a SNIa) emitting light with a known temporal characteristic. A first signal is emitted at some time $t$ by such a source located at a coordinate $x$ and reaches at time $t_0$ an observer located along the $x$ direction at a coordinate $x_0$. A second signal is emitted from the same source at time $t + \Delta t$ and reaches the observer at time $t_0+\Delta t_0$. It can then be shown that, in a perfectly homogenous and isotropic space, we have
	\begin{equation}\label{pep}
		a(t) = a(t_0)\frac{\Delta t}{\Delta t_0}\,.
	\end{equation}
	Here, $a(t)$ is the value of the scale factor at the location of the SNIa that we want to determine. By convention, the value of the current scale factor $a(t_0)$ is set at $1$. Since $\Delta t$ is a characteristic time span supposed to be known, a measure of $\Delta t_0$ allows to deduce $a(t)$ from Eq.\ $(\ref{pep})$. 
	
	In reality, space is not perfectly homogeneous and isotropic, and contains perturbations. As we will see, this can significantly affect the result of the redshift measurement. To show this, let us first write the interval as
	\begin{equation}\label{79}
		ds^2 = g_{\mu\nu}dx^\mu dx^\nu = \left(\overline{g}_{\mu\nu} + \Delta g_{\mu\nu}\right)dx^\mu dx^\nu\,,
	\end{equation}
	where $g_{\mu\nu}$ is the local metric tensor, $\overline{g}_{\mu\nu}$ is the FLRW metric tensor, and $\Delta g_{\mu\nu}$ is the perturbation, i.e., the difference between the real local and the FLRW metric tensors. 
	
	Light follows a null geodesic, so if we consider a light ray travelling along the $x$ direction, we have
	\begin{equation}
		\left(\overline{g}_{tt} + \Delta g_{tt}\right)dt^2 + \left(\overline{g}_{xx}+\Delta g_{xx}\right)dx^2 + 2\Delta g_{xt}dxdt = 0\,.
	\end{equation}
	Solving for $dx/dt$, we find
	\begin{equation}
		\frac{dx}{dt} = \frac{-2\Delta g_{xt} \pm \sqrt{\delta}}{2(\overline{g}_{xx}+\Delta g_{xx})}\,,
	\end{equation}
	where
	\begin{equation}
		\delta = 4(\Delta g_{xt})^2 - 4(\overline{g}_{xx} + \Delta g_{xx})\left(\overline{g}_{tt} + \Delta g_{tt}\right)\,.
	\end{equation}
	We have one solution for a signal travelling in the positive direction, and one solution for a signal travelling in the negative direction. Since space is expected to be isotropic, this equation should on average provide a similar result in magnitude for both signals, but with only a change of sign. This is only possible if on average $\Delta g_{xt}$ vanishes. Then, replacing $\overline{g}_{tt} = -1$ and $\overline{g}_{xx} = a^2$, we get
	\begin{equation}\label{pp1}
		\sqrt{\frac{1-\Delta g_{tt}}{a^2+\Delta g_{xx}}}dt = \pm dx\,.
	\end{equation}
	We now integrate Eq.\ $(\ref{pp1})$ along the $x$ direction. For the signal emitted at $t$, we get
	\begin{equation}
		\int_{t}^{t_0} \sqrt{\frac{1-\Delta g_{tt}}{a^2+\Delta g_{xx}}}dt = \pm\int_{x}^{x_0}dx\,.
	\end{equation}
	Considering the equivalent relation for the second signal we show that
	\begin{equation}\label{ggqq}
		\int_{t}^{t+\Delta t}\sqrt{\frac{1-\Delta g_{tt}}{a^2+\Delta g_{xx}}}dt = \int_{t_0}^{t_0+\Delta t_0}\sqrt{\frac{1-\Delta g_{tt}}{a^2+\Delta g_{xx}}}dt\,.
	\end{equation}
	For a small variation in time, the components of the metric may be considered as constant, and we deduce that
	\begin{eqnarray}\label{ivc}
		\frac{\sqrt{a^2(t)+\Delta g_{xx}(x,t)}}{\Delta t\sqrt{1-\Delta g_{tt}(x,t)}} = \frac{\sqrt{a^2(t_0)+\Delta g_{xx}(x_0,t_0)}}{\Delta t_0\sqrt{1-\Delta g_{tt}(x_0,t_0)}}\,.
	\end{eqnarray}
	In our simplified model, SNIa occur in an overdense region, and we are ourselves located in an overdense region. This means that, using the functions $f$ and $b$ defined above, we have in a typical overdense region
	\begin{eqnarray}
		a^2(t)+\Delta g_{xx}(x,t) &=& b^2(t)\,,
		\\
		1-\Delta g_{tt}(x,t) &=& f^2(t)\,,
	\end{eqnarray}
	at any $t$ and $x$, so Eq.\ $(\ref{ivc})$ can be written as
	\begin{equation}\label{dq}
		\frac{b(t)}{f(t) \Delta t} = \frac{b(t_0)}{f(t_0) \Delta t_0}\,.
	\end{equation}

	When performing redshift measurements, we use Eq.\ $(\ref{pep})$ to determine the scale factor at some time $t$. This thus requires to measure $\Delta t/\Delta t_0$. In practice, however, this is not what we measure. Indeed, the time span measurement performed at our location is carried out in our proper time frame, meaning that we do not measure $\Delta t_0$ but instead $f(t_0)\Delta t_0$. Similarly, the characteristic time span at the source is known in its proper time frame, hence it is equal to $f(t)\Delta t$ instead of $\Delta t$. So, what we measure in practice is not $\Delta t/\Delta t_0$, but instead
	\begin{equation}
		\frac{f(t)\Delta t}{f(t_0)\Delta t_0}\,.
	\end{equation}
	This is the measured value $a_{meas}(t)$ of the scale factor at the SNIa, given that $a(t_0)$ is conventionally set to $1$. From Eq.\ $(\ref{dq})$, we deduce that
	\begin{equation}\label{jj}
		a_{(meas)}(t) = \frac{b(t)}{b(t_0)}\,.
	\end{equation}
	In general, $a_{(meas)}(t)$ will differ from the real value of $a(t)$.	
	
	It is important to notice that this has also consequences on the Hubble constant. This parameter is defined as 
	\begin{equation}
		H_0 = \frac{\dot{a}(t_0)}{a(t_0)}
	\end{equation}
	where $a$ and $\dot{a}$ are measured at the current epoch $t_0$, and it will thus also be affected by the bias. First of all, we need to take into account the fact that we do not measure the real scale factor $a$, but instead a biased factor whose value is given by Eq.\ $(\ref{jj})$. Secondly, we also need to consider that the temporal evolution is determined according to our own proper time, and not to the Cosmological time. As a consequence, the value of the Hubble constant, as measured in practice, corresponds to
	\begin{equation}
		H_{0(meas)} = \frac{1}{f(t_0)}\frac{\dot{a}_{(meas)}(t_0)}{a_{(meas)}(t_0)}\,,
	\end{equation}
	where all variables are evaluated at the current epoch. According to Eq.\ $(\ref{jj})$, we have
	\begin{equation}
		H_{0(meas)} = \frac{1}{f(t_0)}\frac{\dot{b}(t_0)}{b(t_0)}\,.
	\end{equation}
	Using then Eq.\ $(\ref{fq})$, we find
	\begin{equation}
		H_{0(meas)} = \frac{1}{f(t_0)}\frac{V(t_0)}{V_o}\frac{\dot{a}(t_0)}{a(t_0)} = \sqrt{A}H_0\,.
	\end{equation}
	So, the real value of the Hubble constant is related to the measured one by the following relation:
	\begin{equation}\label{H0}
		H_0 = \frac{H_{0(meas)}}{\sqrt{A}}\,.
	\end{equation}
	Since $A>1$, we have $H_0 < H_{0(meas)}$: the real Hubble constant is smaller than the measured one, implying hence also a larger age of the universe.	
	
	\subsection{Effect on luminosity distance measurements}
	
	The luminosity distance $d_L$ is defined as
	\begin{equation}\label{qk}
		d_L^2 = \frac{L}{4\pi F}\,,
	\end{equation}
	where $L$ is the absolute luminosity (supposed to be known) emitted by the source and $F$ is the flux measured by the observer. So, a measure of the flux $F$ completely determines the luminosity distance. Let us therefore examine how the bias affects this measurement. The flux measured by the observer is an amount of energy per unit time and per unit area. This quantity can thus be affected in three ways:
	\begin{enumerate}
		\item The energy that has been emitted by the source has been diluted during its propagation. In a perfectly homogeneous and isotropic space, this dilution has occurred in two ways. Firstly, photons undergo a redshift due to the expansion of the universe, and secondly, photons hit the measurement apparatus less frequently, since two photons emitted a time $\delta t$ apart by the source will be measured with a larger time span by the observer. In the theoretical situation, this double dilution corresponds to a redshift of $(a_0/a)^2$, where $a$ is the scale factor at the location of the source, and $a_0$ is the scale factor at the location of the observer. In the real situation, due to the inhomogeneity, a triple dilution has occurred. First, the a redshift has occurred due to the spatial expansion, corresponding to a dilution factor of $b_0/b$. Secondly, since the first diagonal component of the metric tensor is not constant, a temporal expansion has also occurred, responsible for a loss of energy corresponding to $f/f_0$. And thirdly, as shown by Eq.\ $(\ref{dq})$, in the Cosmological time frame, the ratio between the time span at the source and the one measured by the observer is $b_0f/bf_0$. So, compared with the theoretical situation, in the real situation we expect for the energy that will be measured by the observer a correction factor equal to $(a_0f_0b/afb_0)^2$.
		\item The source emits energy at a rate which is known in its proper time frame, and similarly, the observer measures the flux in its own proper time frame. Therefore, we expect a correction factor of $f/f_0$. Since $f>1$, time progresses at a rate larger than on average through space. This means that during a Cosmological unit time, the source will have more time to emit photons, and will thus emit a larger energy. On the contrary, at the location of the observer, the flux will be smaller, since a same amount of photons will be measured during a larger time span.
		\item Distances are also affected by the metric. The flux is measured by unit area, and since $b>a$, proper distances (and surfaces) in overdense regions are larger than the ones in a perfect homogeneous and isotropic space. This means that, in overdense regions, the flux that will be observed will be diluted over a larger surface, and will hence be smaller. We thus expect a correction factor of $a_0^2/b_0^2$.
	\end{enumerate}
	Considering all these correction factors, the ratio between the measured flux $F_{meas}$ and the one we should have in a theoretical perfectly homogeneous and isotropic space is
	\begin{eqnarray}
		\frac{F_{meas}}{F} &=& \left(\frac{a_0f_0b}{afb_0}\right)^2\left(\frac{f}{f_0}\right)\left(\frac{a_0}{b_0}\right)^2\nonumber
		\\
		&=& \left(\frac{a_0}{b_0}\right)^4\left(\frac{b}{a}\right)^2\frac{f_0}{f}\,.
	\end{eqnarray}
	Consequently, the ratio between the measured luminosity distance $d_{L(meas)}$ and the one we should have in a perfectly homogeneous and isotropic space is such that
	\begin{equation}
		\frac{d_L^2}{d_{L(meas)}^2} = \frac{F_{meas}}{F}\,.
	\end{equation}
	Hence:
	\begin{equation}\label{kk}
		d_{L(meas)} = \left(\frac{b_0}{a_0}\right)^2\left(\frac{a}{b}\right)\sqrt{\frac{f}{f_0}}d_L\,.
	\end{equation}

	\section{Results and discussion}\label{S3}
	
	The whole model relies on one single parameter, i.e., $A$, which has therefore to be quantitatively estimated. This parameter represents the current value of $V/V_o$. According to several references, void regions occupy now about $80\%$ of space. For example, $\cite{Cautun}$ estimates that void regions represent currently $77\%$ of space, while \cite{Falck} obtains values slightly above $80\%$. According to \cite{Tavasoli}, void regions occupy more than $80\%$ of the volume of the observable universe. Even if it should be stressed that those values depend on how void regions were defined, this value of $80\%$ has to be considered as an order of magnitude of the volume occupied by void regions. As a consequence, overdense regions occupy currently about $20\%$ of space, implying that $A \simeq 5$. 
	
	A parametric study has been performed with the proposed model to select the value of $A$ that fits the measurements in the best way. This has led to fix $A = 5.13$, corresponding to a volume of overdense regions occupying about $19.5\%$ of space. This is in agreement with the order of magnitude mentioned above. This is the value that has thus been used to establish the results that follow.
	
\begin{figure}
		\centering\includegraphics[width=12cm]{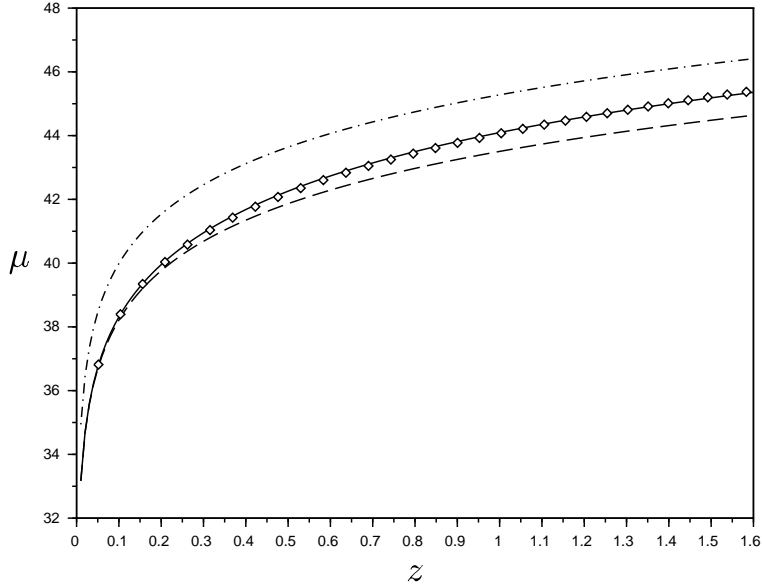}
		\caption{Distance modulus versus redshift. Dashed line: $\Omega_{m,0} = 1$, $\Omega_{\Lambda,0} = 0$ and $H_0 = 70\ km/s/Mpc$; solid line: $\Omega_{m,0} = 0.3$, $\Omega_{\Lambda,0} = 0.7$ and $H_0 = 70\ km/s/Mpc$; dash-dot line: $\Omega_{m,0} = 1$, $\Omega_{\Lambda,0} = 0$ and $H_0 = 31.3\ km/s/Mpc$; diamonds: theoretical prediction of the measured relation.}\label{Fig1}
	\end{figure}

	We start from the Hubble constant deduced from SNIa measurements, which corresponds to a value of about $70\ km/s/Mpc$, see for example $\cite{Riess2}$. In Figure $\ref{Fig1}$, the dashed line represents the distance modulus $\mu$ versus redshift relation for a universe with $\Omega_{m,0} = 1$ and $\Omega_{\Lambda,0} = 0$, with such a Hubble constant. This relation is however not the one that is observed. Measurements indeed show an accelerated expansion, for which $\Omega_{m,0} = 0.3$ and $\Omega_{\Lambda,0} = 0.7$. The corresponding distance modulus versus redshift relation is plotted in Figure $\ref{Fig1}$ with a solid line.
			
	As explained above, the Hubble constant has been measured in our proper time, which differs from the Cosmological time. To find the Hubble constant as it should be determined in a perfectly homogeneous and isotropic space, we use Eq.\ $(\ref{H0})$, and we find that the the real value is $H_0 = 31.3\ km/s/Mpc$, hence quite smaller than the measured one. In Figure $\ref{Fig1}$, the dash-dot line represents the distance modulus versus redshift relation for a universe with $\Omega_m = 1$ and $\Omega_\Lambda = 0$, with this latter Hubble constant. This is the relation that should be obtained if there was no bias in the measurements. However, due to the bias, a different relation is observed. To determine this one, we use Eq.\ $(\ref{jj})$ and $(\ref{kk})$ to deduce how luminosity distance and redshift measurements have been affected by the bias. The resulting distance modulus versus redshift relation is plotted in Figure $\ref{Fig1}$ with diamonds.
	
\begin{figure}[h]
		\centering\includegraphics[width=12cm]{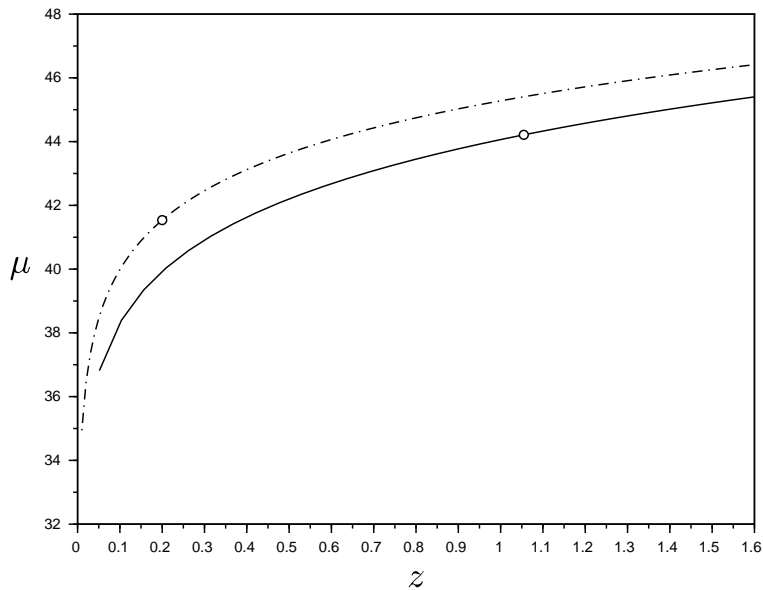}
		\caption{Illustration of the effect of the bias for one particular measurement on the distance modulus versus redshift relation.}\label{Fig2}
	\end{figure}

	An excellent correspondence between the measured relation and the predicted one can be observed. To illustrate the effect of the bias on a specific measurement, we plot in Figure $\ref{Fig2}$ two circles: the one on the upper curve is the one that should be measured in a perfectly homogeneous and isotropic space. But due to the perturbations in the metric at the source as well at the location of the observer, redshift and luminosity distance measurements are strongly affected, and send the result to the circle located on the lower curve, far away from the real situation.
		
\begin{figure}
		\centering\includegraphics[width=12cm]{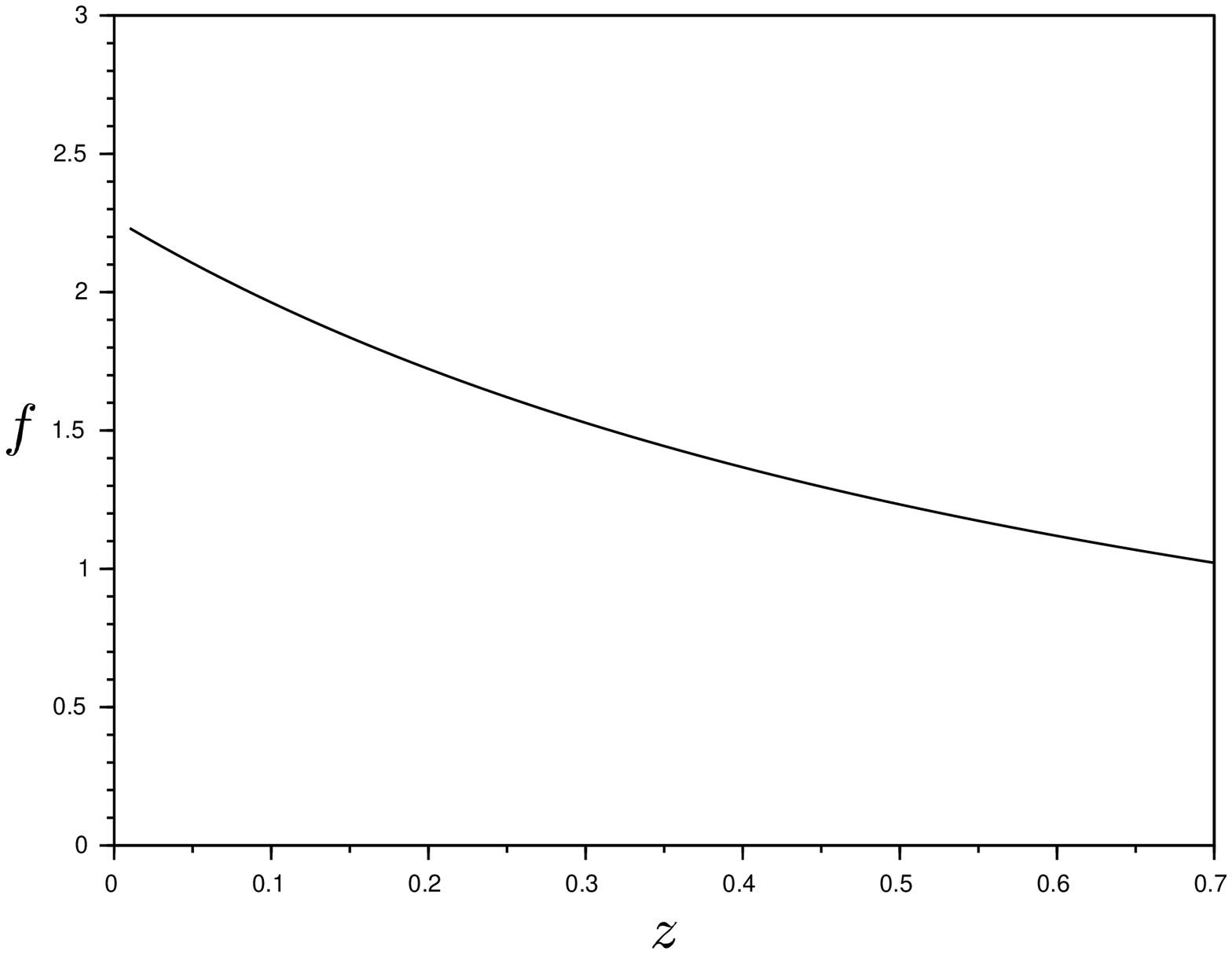}
		\caption{Evolution of $f$ in function of $z$.}\label{Fig3}
	\end{figure}

	In order to grasp to effect of the metric, we plot in Figure $\ref{Fig3}$ the evolution of $f$. In a perfectly homogeneous and isotropic space, $f$ is constant over time and equal to $1$. In overdense regions, however, $f$ continuously increases and so strongly diverges from what it should be on average over space. At our current epoch, it has already reached a value of about $2.25$, meaning that time progresses at a rate $2.25$ times larger than on average through space.
	
\begin{figure}
		\centering\includegraphics[width=12cm]{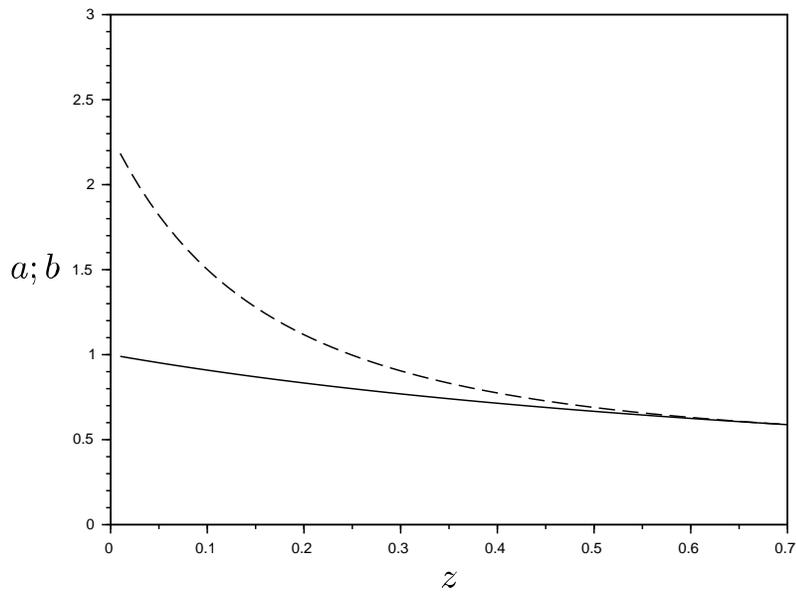}
		\caption{Evolution of $a$ (solid line) and $b$ (dashed line) in function of $z$.}\label{Fig4}
	\end{figure}

	Also, Figure $\ref{Fig4}$ represents the evolution of $b$ in function of the redshift $z$ (dashed line), and compares it with the evolution of the scale factor $a$ of the average space (solid line). Whereas, by convention, we have fixed $a=1$ at our current epoch, we observe that $b$ has reached a value of about $2.3$, meaning that space in overdense regions has expanded quite faster.
	
	We should finally stress that in order to develop the model we have made a strong assumption on the volume of overdense regions. Considering that matter has immediately reached a gravitational bound state and hence that overdense regions have a constant volume is an extreme situation, and this could quantitatively affect the way how $f$ and $b$ evolve over time. The advantage of this assumption is that the model could completely be developed analytically. The drawback is that, even if the model reproduces satisfactorily the measured distance modulus versus redshift relation, it is not the more realistic one. In particular, we could expect that some values determined from the model (such as the real Hubble constant for example) could be affected in some way.	A more realistic situation would require to consider a given evolution of $V/V_0$, starting from a value of $1$ at the origin of time, and tending progressively to be proportional to $a^3$. This would however require additional parameters to be included into the model, leading to a more complex approach. Moreover, establishing a more realistic evolution of $V/V_0$ can be quite challenging also. 
	
	Nevertheless, even with the strong simplifying assumption, the results obtained with the model developed in this article are promising, but they should be considered more qualitatively than quantitatively. The model provides an insight of the phenomena that could explain the apparent accelerated expansion of the universe as evidenced by SNIa measurements, without needing to assume the existence of dark energy. It is finally important to highlight that the consequences of the perturbation in the metric in overdense regions exceed the case of the SNIa measurements. All kinds of measurements performed from our specific location use apparatuses and theories that generally rely on temporal and spatial concepts, and could thus be biased in a similar way. Results made on astrophysical phenomena should thus be interpreted cautiously. This is in particular true for BAO and cosmic microwave background measurements.
	
	
	\section{Conclusion}
	
	Admitting that SNIa measurements are affected by a bias, related to the fact that SNIa occur only in overdense regions, we have developed a model to investigate the effect of this bias on the measurement results. This model relies on one single parameter, i.e., the part of space that is occupied by overdense regions. The model considers two distinct regions, namely overdense and underdense regions, and assumes that those regions can be described by average metric and stress-energy tensors. We then focussed on establishing the average metric tensor in overdense regions. This metric tensor presents a scale factor differing from the one of the average space, but also a function that describes the rate at which time progresses in overdense regions. This rate may indeed differ from the one of the average space. Considering the metric tensor in overdense region, the effect of the bias on the results of redshift and luminosity measurements has then been examined. Such measurement indeed imply temporal and spatial concepts, and are thus affected by the perturbation existing in the local metric tensor at the source as well as at the location of the observer. Due to the different scale factor and the different rate at which time progresses in overdense regions, it has been shown that the results of those measurements are strongly affected. Assuming a void fraction of space of about $80\%$, similar to the values found in the literature, it was shown that the model predicts a distance modulus versus redshift relation in perfect agreement with the one established with the SNIa probe (corresponding to $\Omega_{m,0} = 0.3$ and $\Omega_{\Lambda,0} = 0.7$). According to the proposed model, the apparent accelerated expansion of the universe can be explained as a measurement artefact, which thus does not require to assume the existence of some kind of dark energy.



	
\end{document}